\documentclass[preprint2]{aastex}

\slugcomment{to be submitted to The Astrophysical Journal}

\shorttitle{Planetary Nebula Distance Scale}
\shortauthors{Stanghellini et~al.}
  
\begin{document}
 
\title{The Magellanic Cloud Calibration of the Galactic Planetary Nebula Distance Scale}
\author{Letizia Stanghellini}
\affil{National Optical Astronomy Observatory, 950 N. Cherry Av.,
Tucson, AZ  85719}
\email{lstanghellini@noao.edu}

\author{Richard A. Shaw}
\affil{National Optical Astronomy Observatory, 950 N. Cherry Av.,
Tucson, AZ  85719}
\email{shaw@noao.edu}

\author{Eva Villaver\altaffilmark{1}}
\affil{Space Telescope Science Institute, 
3800 San Martin Drive, Baltimore, MD 21218}
\email{villaver@stsci.edu}

\altaffiltext{1}{Affiliated with the Hubble Space Telescope Division
of the European Space Agency}

\begin{abstract}
Galactic planetary nebula (PN) distances are derived, except in a small
number of cases, through the calibration of statistical properties of
PNe. Such calibrations are limited by the accuracy of individual PN
distances which are obtained with several non-homogeneous methods, each
carrying its own set of liabilities. In this paper we use the physical
properties of the PNe in the Magellanic Clouds, and their accurately
known distances, to recalibrate the Shklovsky/Daub
distance technique. Our
new calibration is very similar (within 1\%) of
the commonly used distance scale by Cahn et al. (1992), although there
are important differences. We find that neither distance scale works well
for PNe with classic ("butterfly") bipolar morphology, and while the
radiation bounded PN sequences in both the Galactic and the Magellanic
Cloud calibration have similar slopes, the transition from optically
thick to optically thin appears to occur at higher surface brightness and smaller
size than that adopted by Cahn et al. The
dispersion in the determination of the scale factor suggests that PN
distances derived by this method are uncertain by at least 30\%, and
that this dispersion cannot be reduced significantly by using better
calibrators. We present a catalog of Galactic PN distances
using our re-calibration which can be used for future applications, and
compare the best individual Galactic PN distances to our new and several other
distance scales, both in the literature and newly recalibrated by us, finding that our
scale is the most reliable to date.

\end{abstract}


\section{Introduction}

The uncertainty associated with distances measurements of Galactic
planetary nebulae (PNe) is 
a major obstacle to the advancement of PN research. Only $\sim$40 Galactic PNe 
have distances that have been determined individually with reasonable accuracy. 
Distances to Galactic PNe can be determined individually in various
ways, including cluster membership (Chen et al.~2003, CHW03; Alves et al.~2000, ABL00),
by measuring the rate of their expansion (e.~g., Liller \& Liller 1968, LL68; Hajian et al. 1995, HTB95), 
by the reddening
method (e.~g., Gathier et al.~1986, GPP96; Kaler \& Lutz 1985, KL85), and by measuring their 
spectroscopic parallax 
(Ciardullo et al. 1999, C99) or trigonometric parallax  (Harris et al. 2007, Hea07). 

For the remaining $>$1800 Galactic PNe (Acker et al.~1992) one has to rely on
statistical distance scales,  
whose calibrations are based on the reliability of the individually known PN distances, 
and the validity of a general correlation that links the distance-dependent to
the distance-independent physical properties of PNe.   
The Cahn, Kaler, \& Stanghellini (1992, CKS) distance scale of
Galactic PNe is based on an attempt by Daub (1982, D82) to improve  
Shklovsky's distance scale (Shklovsky 1956a, 1956b) for optically thick nebulae.
Shklovsky's distance scale assumes that all PNe have equal (observed) ionized mass. 
D82 assumed that Shklovsky's constant mass approach was still valid, but only for 
those PNe that are optically thin to the Lymann continuum radiation emitted by
the central stars (density 
bounded). For the optically thick (radiation bounded) PNe, D82 based the
distance scale on a calibration 
of an ionized mass versus surface brightness relation. CKS improved
D82's calibration with the 
use of a larger number of calibrators (PNe with known individual distance), and calculated
the statistical distances 
to 778 Galactic PNe.

Since its publication, distances from the CKS catalog have been used
preferentially and widely in the literature. Other statistical methods that have
been commonly used include those by Maciel (1984), Zhang (1995, Z95), van de Steene \& Zijlstra 
(1995, vdSZ), Schneider \& Buckley (1996, SB96) and Bensby \& Lundstr{\"o}m (2001, BL01). All these distance 
scales rely on a set of Galactic calibrators whose distances are mostly
derived from reddening or expansion properties, or from the assumption of 
Galactic Bulge membership, with all the consequent uncertainties. 
With the publication over
the past decade of critical physical parameters for a large sample of
Magellanic Cloud PNe (Shaw et al. 2001, 2006; Stanghellini et al. 2002,
2003), including highly accurate H$\beta$ fluxes, physical dimensions,
morphologies, and extinction constants, we have the opportunity to assess
and improve the distance scale for Galactic PNe. In this paper we take advantage of the 
wealth of Magellanic Cloud PN data to re-calibrate the CKS
distance scale, as well as other distance scales for comparison. Homogeneously determined 
photometric radii from {\it HST} of a PN sample with low Galactic reddening are the best way to determine 
any relation that involves apparent diameters.
Furthermore, the relatively recent publication of trigonometric and spectroscopic parallax and cluster membership 
distances to Galactic PNe allow us to 
test with unprecedented reliability our own and other distance scales.

The construction of any statistical distance scale for PNe is composed of three fundamental steps: 
the selection of a \textit{method} that has some physical or empirical basis, the selection of a set of 
\textit{calibrator PNe}, for which distances have been determined by some independent means, and an analysis 
of the \textit{applicability} of the calibration to a wide variety of PNe. Until now it has been difficult 
to compare the viability various methods, since their calibrations and applications have varied so widely. 
In \S2 we describe in detail the Shklovsky/Daub/CKS distance scale and the superiority of the Magellanic 
Cloud PNe as calibrators. We also derive our new calibration of this method and assess its inherent uncertainties. 
In \S3 we discuss the physical underpinning of the CKS distance method in light of recent advances in modeling the 
evolution of PNe. In \S4 we take a closer look at the viability of various methods for determining independent 
distances to Galactic PNe (i.e., the set that had previously been used as calibrators), and discuss the 
applicability of the Magellanic Cloud distance scale to Galactic PNe. In \S5 we recalibrate the most used statistical 
distance methods with the Magellanic Cloud PNe, and then compare the accuracy of these methods to one another using 
the best independently determined distances to Galactic PNe. We conclude in \S6 with our final prescription and 
recommendation for determining statistical distances to Galactic PNe.

\section{The Magellanic Cloud 
  Calibration and the New PN Distance Catalog} 

The CKS statistical distance scale is based on the calibration of the relation 
between D82's {\it ionized mass} 

$$\mu= (2.266 \times 10^{-21} D^5 \theta^3 F)^{1/2} \eqno(1)$$

and the {\it optical thickness parameter} 
$$\tau = {\rm log} {4 \theta^2\over F},\eqno(2)$$
 
where D is the distance to the PN in parsecs, $\theta$ is the nebular radius 
in arcsec, and F is the nebular flux at 5 GHz. 
The parameter $\mu$ increases as the ionization front expands into the
nebula. Once a PN becomes density bounded, $\mu$ remains constant
for the rest of the observable PN lifetime.

By calculating $\mu$ and $\tau$ for several PNe with known distances, dimensions, and fluxes, CKS
derived the $\mu$--$\tau$ relation:
$${\rm log}~\mu = \tau-4, ~\tau < 3.13 \eqno(3a)$$
$${\rm log}~\mu = -0.87, ~\tau > 3.13, \eqno(3b)$$

where Eq. 3a holds for PNe of high surface brightness, and Eq. 3b for PNe with low surface brightness.

The calibration of the above distance scale was based upon 19 Galactic PNe with independent distances
with comparatively poor accuracy.
At the time when the CKS paper was written there were hardly any Magellanic
Cloud PNe with accurately measured diameters, and the distances to the Magellanic
Clouds were also quite uncertain. We can now re-calibrate the distance scale using the 
nebular parameters relative to the LMC and SMC PNe observed by us with the {\it Hubble Space
Telescope (HST)} (Shaw et al.~2001, 2006; Stanghellini et al.~2002, 2003). 
In order to determine $\tau$ and $\mu$ for Magellanic Cloud PNe we use a transformation 
between the 5 GHz and the H$\beta$ fluxes (Eq. 6 in CKS), since radio fluxes are not
available for Magellanic Cloud PNe. All other parameters are available in our 
{\it HST} paper series. Note that we use the photometric radius as the proper measure of the
nebular dimension, which is defined as the radius that includes 85\% of
the flux in a monochromatic emission line.

We have adopted a distance to
the LMC  of 50.6 {\rm kpc} (Freedman et al.~2001; Mould et al.~2000), which is accurate to
$\sim$10\%  (Benedict et al.~2002).
The variation in the adopted distance
when applied to individual objects can be easily estimated  given that 
the three dimensional  structure of the LMC has been well
established (Freeman, Illingworth, \& Oemler 1983; van
der Marel \& Cioni 2001). The LMC can be 
considered a flattened disk with a tilt of the LMC plane to the plane of the
sky of 34$^\circ$ (van der Marel \& Cioni 2001).  Freeman, Illingworth, \&
Oemler (1983) derived a scale height of 500  
{\rm pc} for an old disk population. The scale 
height of young objects is between 100 to 300 {\rm pc}
(Feast~1989). Using the scale height of an old disk population the  3D
structure of the LMC introduces a variation in the adopted 
distance smaller than 1\% from object to object and therefore has been neglected in the
calibration.

For the  SMC we have used a distance of  58.3  {\rm kpc} (Westerlund 1997).
The accuracy of this distance is not as well established as for the LMC. 
Moreover, the SMC is irregular with a large intrinsic line
of sight depth (between 6 and 12 {\rm kpc}: Crowl et al. 2001) which varies
with the location within the galaxy. We have estimated an average line of sight
depth of 5 {\rm kpc} for the PNe in our sample by combining the span of the 
positions (400 {\rm pc} in right
ascension and 2 {\rm kpc} in declination)  of the PNe with respect to the optical center
of the SMC with the dispersion in the distance to the SMC derived by Crowl
et al. (2001) using SMC clusters positions. 
The distance uncertainty introduced by this
depth in the SMC is roughly 9\%, still low to significantly
affect the result but one order of magnitude larger than the one obtained for the LMC. In this
respect we consider  LMC PNe to be better calibrators than the SMC 
PN for the distance scale.

In Figure 1 we show the LMC PNe on the log~$\mu$ - $\tau$ plane. 
We have calculated $\tau$ and log~$\mu$ as explained above, and assumed
D$_{\rm LMC}$=50.6 kpc. In the Figure we plot the different morphological types with different 
symbols, following the classification in Shaw et al.~(2001, 2006).
To guide the eye we have plotted, on the figure, the 
Galactic distance scale fit from CKS (solid line).
The optically thick
sequence of LMC PNe is very tight for $\tau<$2.1, and most LMC PNe are optically thin for
$\tau>$2.1. The fitted value of the function for optically thin LMC PNe
is almost  
identical to that of Galactic PNe {\it if we exclude bipolar planetary
nebulae}. 
The broken
line in Figure 1 corresponds to the Magellanic Cloud fit of the optically thick sequence of LMC PNe 
(see Eq. 4a and 4b below).
Similarly, in Figure 2 we show the same plot of Figure 1, but for SMC PNe. The morphology and sizes of the nebulae are 
from Stanghellini et al.~(2003). 
Even with the scarcity of data points, the thick PN sequence is well defined by SMC PNe,
and it is identical to that of the LMC PNe.

The observed ionized masses of bipolar PNe in both Figures 1 and 2 appear mostly well above the constant ionized
mass line. The parameters
$\mu$ and $\tau$ have been calculated with the photometric radii of the PNe,
that can be very different from the isophotal radii in the case
of PNe with large lobes. Furthermore, bipolar PNe might be optically thick for most of their
observed lifetime (Villaver et al. 2002a), and thus are not the ideal calibrators 
for the optically thin PNe branch
of the log $\mu$ -- $\tau$ relation. 
In deriving the distance scale based on Magellanic Cloud PNe we thus exclude 
PNe with bipolar morphology. This leave us with 70 Magellanic Cloud calibrators, 
a very large number of PNe with individual distances when compared to the 19 
calibrators in CKS. In Figure 3 we show the 
Magellanic Cloud calibration of the PN distance scale, where open symbols are LMC PNe and filled symbols are SMC PNe,
and where we exclude bipolar PNe. Note that we have assumed that the ionized mass for optically thin PNe is constant,
as in D82 and CKS.

The fit to the distance scale based on the Magellanic Cloud PNe (this paper,
hereafter SSV) is:

$${\rm log}~\mu = 1.21~ \tau -3.39, ~\tau < 2.1 \eqno(4a)$$
$${\rm log}~\mu = -0.86, ~\tau > 2.1 \eqno(4b)$$

The solid line in Figure 3 shows this relation. The separation between optically thick and thin
PNe is very obvious from the figure, and the optically thick sequence is much better defined here than
in CKS, thanks to the use of the best calibrators available now. 
The optically thick sequence has been derived by least-square fit, and has correlation coefficient R$_{\rm xy}$=0.8.
The 
optically thin sequence is determined by the average of the log $\mu$ for $\tau>2.1$.
Using another estimate of the central tendency will change the horizontal scale by less than
5$\%$, which is well within the uncertainty. Furthermore, if we were to fit
the data points of Figure 3 with just one line for all $\tau$ we would have a very poor correlation (R$_{\rm xy}$=0.14), which reinforce the 
evolutionary scheme of optically thick to thin PNe, proposed by D82 to improve Shklovsky's method. 

By examining Figure 3 we infer that: 
(1) Our analysis allows us to confirm the 
CKS distance scale for optically thin PNe; (2) the optically
thick sequence 
is very well defined by the Magellanic Cloud PNe and it is different from that of CKS;
(3) the new statistical distance for optically thin PNe increases slightly the assumed ionized mass, such that 
distances for optically thin nebulae are tipically
1$\%$ larger compared to those computed using the CKS calibration.
(4) bipolar PNe do not follow the empirical relation, and their ionized mass actually increases steadily 
with $\tau$,
confirming that they stay in the ionization bound state for much longer than PNe with other morphological types.
The probable reason why the bipolar PN relation does not flatten out
for $\tau>$2.1 is because they are the progeny of the more
massive stars and they are expected to remain optically thick (given a
combination of the large circumstellar densities and fast evolution of the
central star).

By using the SSV distance scale we calculated the statistical distances to all non-bipolar PNe in the LMC and 
the SMC. We obtain distributions that are nicely narrow, with mean values (and dispersions),
D$_{\rm LMC}$=50.0$\pm$7.5 kpc and D$_{\rm SMC}$=57.5$\pm$5.5 kpc,
that are within 1$\%$ of the distances to the Magellanic Clouds.

We applied our new distance scale to the large sample of Galactic PNe in
the original CKS catalog, and present the revised distances in Table 1.
Column (1) gives the usual name as in CKS, column (2) gives the calculated $\tau$, columns
(3) and (4) give the angular radius and the flux used in the calculation, and column (5)
gives the distance to the PNe. Note that the fluxes in (4) are the 5 GHz fluxes from CKS when available,
or their H$\beta$ equivalents.

\section{The Physics of the Statistical Distance Scale}

As CKS pointed out, the assumption of constant ionized mass for optically
thin PNe (or that it can be computed for optically thick PNe with a
one-parameter model) would seem to be a doubtful proposition since the
progenitor stars vary in mass by nearly an order of magnitude. CKS minimized
the significance of the variation in ionized mass by pointing out that
distances so derived depend only on the square-root of the assumed mass.
One might also expect that the ionized
mass would be fairly directly correlated with the progenitor mass. However,
hydrodynamical models of the co-evolving PN and central star by Villaver et
al. (2002a) show that the decline of gas density with radius is generally
quite steep (except within the bright inner shell of gas) over a wide range
of progenitor masses and during the entire visible lifetime of the nebula.
The implication is that, for optically thin nebulae, the bulk of the mass
exists in the faint, low-density, outer halo. Since the volume emissivity
of recombination lines is proportional to the square of the gas density,
the massive nebular halo contributes very little to the observed emission.
Most published values for PN masses assume a constant density for the gas,
one that is only representative of the bright inner shell, leaving the bulk
of the PN mass unaccounted for. In part for these reasons, ionized masses
derived in this way reflect only a modest fraction of the total mass of the
nebula, such that the assumption of a constant mass is sufficiently accurate
to render the Shklovsky distance method useful.

We have shown that the distance method of CKS is empirically sound, and derived
the scale factor for optically
thin PNe to that from  observations of
Magellanic Cloud PNe. It is important to note the significance of the
{\it dispersion} in the PN masses (expressed in the $\mu$ parameter)
about the mean in the calibration shown in Figure 3. The $1-\sigma$
deviation about the mean value is 0.28, which translates to a
corresponding uncertainty in the distance of about 30\%. We regard this
value as a rough estimate of the minimum uncertainty that may be
associated with the distance to an individual PN derived using
this methodology. It is important to note that the uncertainty in the
distance scale {\it cannot} be reduced with improved calibrator nebulae,
since the distance uncertainty is of order 10\% (i.e., of order the size
of the symbols in Figure 3). The scatter in the data results from genuine
variations in the ionized masses of the calibrator nebulae, and quantifies
the fundamental limitation in this technique.

The new PN distance scale (SSV) is very similar to
that of CKS, with the exception of the transition between optically thick and optically think stages. 
From Eq.~4b, the definition of $\mu$, and the relation $D=206265 ~ R_{\rm PN}/\theta$ 
(where R$_{\rm PN}$ is the linear nebular radius in pc) we can determine the radius at which the PN becomes optically
thin. For  $\tau$ =2.1 we obtain $R_{\rm PN}\sim$0.06 pc. The same calculation to determine the PN radius at which the thick to thin 
transition occurs by using Eq. 3b and $\tau$=3.13 gives
$R_{\rm PN}\sim0.09$ pc. The uncertainty in the determination of $\tau$ at transition, and thus of 
R$_{\rm PN}$, depends on the scatter of the ionized mass calibrators used in CKS. The new calibration is much more reliable.

The metallicities of the LMC and SMC are, on average, of the order of
half and a quarter that of the solar mix respectively
(Russell \& Bessell 1989; Russell \& Dopita 1990).  
The AGB wind is likely to be dust driven, therefore it has a strong dependency 
on metallicity. It is then 
expected that LMC and SMC stars with dust-driven winds 
lose smaller amounts of matter (Winters et al. 2000) during the AGB phase than their
Galactic counterparts.  The mass-loss history
during the AGB determines the circumstellar density structure that will
eventually constitute the PN shell (Villaver et al. 2002b). A reduced mass-loss rate
during the AGB has the effect of decreasing the density of the
circumstellar envelope prior of PN formation.  

Furthermore, after the envelope is ejected, the remnant central star leaves the AGB and its effective temperature
increases. The stellar remnant becomes a strong emitter of ionizing photons,
responsible for ionizing
the nebula. The mechanism
that drives the wind during the central star phase (with velocities a few orders of
magnitude higher than that experienced during the AGB phase) is the transfer
of photon momentum to the gas through absorption by strong resonance lines
(Pauldrach et al. 1988). The efficiency of this mechanism depends on metallicity,
thus it is expected to be less efficient in Magellanic Cloud central stars than in the Galactic ones,
with correspondingly lower escape velocities for the winds, and a decreased efficiency in 
shell snow-plow.  
  
As has been shown by Villaver et al. (2002a),
the propagation of the ionization front determines the density structure of
the nebula early in its evolution, while
the pressure provided by the hot bubble has no effect at this stage. 
The propagation velocity of the Str\"omgren radius, which 
ultimately determines the transition from the optically thick to the optically thin stages,
depends mainly on the ionizing flux 
from the star and on the
density of the neutral gas. Given the dependency of the AGB mass-loss rates 
on metallicity, the ionization front will 
encounter a lower neutral density structure in Magellanic Cloud PNe than in Galactic PNe. 
This would tend to make the transition from optically thick to thin at a smaller radius in Magellanic 
Cloud PNe than in Galactic PNe. 
The fact that our Magellanic Cloud calibration of the CKS scale occurs at smaller radii than that derived by CKS
is probably coincidence. On the other hand, if we really could
 determine empirically the transition radius as a function of metallicity, we would expect two different 
thick sequences for the SMC and the LMC PNe, given their different metallicity,
and yet the sequences are almost identical (Figs. 1 and 2). 
That is, we do not see the effects of metallicity on our distance scale, and that is applicable to
Galactic PNe as well. We discuss below ($\S$4) 
how the newly derived distances
match extremely well with the best individual distances to Galactic PNe independently of metallicity.

\section{Comparison of our Distance Scale to Individual Galactic PN Distances}

We have assessed that our new calibration of the PN distance scale 
is very similar to that of CKS, but with a revision in the transition between
the radiation bounded and the density bounded stages. The comparison between the CKS and the SSV scales
suffers from the fact that part of the CKS calibrators are obsolete, and that new Galactic calibrators have 
become available. It is worthwhile to compare the SSV scale with the best available individual distances to
Galactic PNe to date before we confirm the validity of the new calibration.

In Table 2 we give the best set of individual Galactic PN distances available to date. 
Column (1) gives the common name; columns (2) and (3) give the best individual distance and, where available,
its
uncertainty; columns (4) and (5) give respectively the 
statistical distances for the same PNe from CKS and the SSV; columns (6) and (7) give the distance
determination method (CM for cluster membership, 
P for parallax, E for expansion, R for reddening, see explanations below) and 
its reference. We have selected a sample of individual Galactic PN distances based on the literature, and 
whose statistical distances have been calculated by CKS and can be derived for the SSV calibration as well.

The best methods to get individual PN distances are (1) trigonometric parallax, (2) the use of a spectroscopic companion of the PN central
star, which allows to derive the spectroscopic parallax, and (3) the membership of the PN in an open or globular cluster.
Apart from trigonometric parallaxes, that are applicable only for nearby PNe, the distance to the PN is that of a companion or a cluster, whose uncertainties are typically much
lower than those related to other methods for PN distances. 
In the past decade there have been two major studies of PN parallaxes. C99 used {\it HST} imaging to
determine central star companions of a PN sample, obtaining ten probable associations and the relative spectroscopic parallaxes. We list all of these 
in Table 2, except for A~31, where 
only a lower limit to the distance is given, and A~33 and K~1-27, whose distances seem to be controversial 
in C99. Hea07 published trigonometric parallaxes of several Galactic PNe. Following the discussion in Hea07, 
we include all their final determinations in Table 2, including the uncertainties.
Planetary nebulae whose distances have been derived through cluster membership are the PN in Ps~1, whose distances has been recalculated
by ABL00, and that in the open cluster NGC~2818, whose distance has been estimated by CHW03.
It is worth noting that Mermilliod et al. (2001) found that the radial velocity of the NGC~2818 PN is slightly lower than
that of the cluster, making its membership marginally questionable. 

Since CKS was published there have been other PNe observed in clusters,
including JaFu1 and JaFu2 (Jacoby et al.~1997), and a PN in M~22 (Monaco et al.~2004), but 
their distances are not included in Table 2 either because their cluster membership is not
definitive or because their nature is still uncertain, as described in detail in the discovery papers.

An alternative method for PN distances is the determination of the secular PN expansion, a method that had
its renaissance with the use of the accurate relative astrometry afforded by the {\it HST}. In this category we found distances to several
PNe by Hajian et al. (1993, 1995, 1996; HBT93, HBT95, HT96), Palen et al. (2002,  Pea02), Gomez et al. (1993, GRM93),
and also the work by LL68. Among the distances determined by expansion we have only listed in Table 2 those
deemed reliable by the authors listed above. In particular, in Pea02 there are several distances determined by different expansion algorithms, and
if the results are very different by different methods for the same PN we have excluded them. Uncertainties in expansion distances, when available,
are much higher than those of parallaxes or cluster membership, and the method is intrinsically less reliable, given the impossibility of
following the PN acceleration history, and the modeling difficulty given unknown process as such as differential mass-loss.

Finally, a very rough individual distance can be derived in some cases by studying the reddening patches around the PN and then building 
reddening-distance plots for the known stars surrounding the PN. This method, although providing several data-points in the literature, is the most uncertain
given the inhomogeneity of the Galactic ISM. GPP96 derived reddening distances for several
PNe, and we used in Table 2
only the reliable ones, as deemed by the authors. We have excluded NGC~2346, since the scatter in its distance-reddening plot is overly large. 
KL85 also published several PN distances by the same method, and we
included their results in Table 2. 

In Figure 4 we plot the data of Table 2 on the $\tau$ -- log~$\mu$ plane, drawing also the CKS and the SSV distance scales (note that the 
scales of Figures 3 and 4 are different). It is interesting to note that the best data points, those of the PNe whose distances
have been determined via parallax or cluster membership, follow very well the SSV calibration, and they are less compatible with that of CKS.
Naturally, the CKS calibration was based principally on reddening and expansion distances, since very few parallaxes were available at that time,
and we can see that the thinning sequence determined by the data points relative to expansion or reddening distances is compatible with
the CKS calibration (but these individual distances have much lower reliability than the parallaxes and cluster memberships represented by the filled symbols). 
While the SSV seems to be the best statistical scale to be used for Galactic PNe, its preference, for
optically thick PNe, over the CKS scale is based only on one
data point (the parallax at $\tau<$2.1). But let us recall here that the filled symbols {\it are not} the calibrators of the SSV scale, rather 
they are the Galactic PNe with best individual distances to date, used for comparison, while the calibration is based on $\sim$70 data points whose errorbars would be
smaller than the symbols.

In Figure 5 we show the direct comparison of the individual PN distances and those from the SSV calibration, where the correspondence of the
parallax and cluster membership individual distances with the SSV distances is remarkable. It is worth noting that the two lower-left 
filled circles, those for which the parallax and statistical distances do not coincide within 30$\%$, are A~7 and A~31; both are very large nebulae,
whose diameters are larger than the radio beam used to detect the 5 GHz flux (Milne 1979), and whose flux densities are deemed to be uncertain. 
In Figure 6 we show the distribution of the relative differences of the SSV and individual distances versus $\tau$ for the four methods of deriving individual distances.
The thin vertical lines represent the $\tau$=2.1 and $\tau$=3.1, i.e., the thick-to-thin PN transition for the CKS and SSV scales. 
We could conclude that the SSV scale fails to reproduce the individual distances for PNe around the transition between thick to
thin, but this failure seems to pertain only to the comparison with expansion and reddening distances, and 
it does not occur for the comparison with parallax and cluster membership distances. 

\section{Comparison of Statistical Distance Scales}

We compare the relative merits of the new SSV scale, calibrated on Magellanic Cloud PNe, 
in relation to other distance scales in the literature. We compare the statistical distances from different methods 
to individual Galactic PN distances, by using only the best individual PN distances of Table 2, those from parallax and 
cluster membership. We also calculate the distances to all LMC and SMC PNe using the statistical methods, then 
we compare the resulting averages with the actual distances to the Clouds. It is worth recalling that
all old scales have been calibrated with Galactic or Bulge PNe, thus we expect a lower reproducibility of the
Magellanic Cloud distances. 

For all scales we give in Table 3: in column (1)  the reference, in column (2) the statistical method, in column (3) 
the correlation coefficient between statistical and
individual PN distances, in column (4) the mean relative difference between the statistical and individual distances, in column (5) the
relative difference between the median distances to Large Magellanic Cloud PNe and the actual LMC distance, in column (6) the same relative difference, but for
the SMC PNe. Statistical distance scales in the literature use in this comparison are those by CKS, vdSZ, Z95, SB96, and BL01. 

The statistical scheme that best compares with the best individual distances is the SSV scale, with higher 
(R$_{\rm xy}$=0.99) correlation and lower median difference between statistical and individual distances than any other scale, and
best reproduces the Magellanic Cloud distances.
This is hardly a surprise, since for the first time it was possible to calibrate a distance scale with absolute calibrators from the
Magellanic Clouds. 
By comparison, the correlation coefficients between the distances from the CKS, vdSZ, Z95, SB96, and BL01 scales are always lower, and the 
median of the relative differences are higher. 

We also want to test whether PN distances derived with other distance scales, if recalibrated with the Magellanic Cloud PNe,
would compare better than the SSV scale to the best individual distances of Galactic PNe. First, we consider
the relation between the brightness temperature and the linear nebular radius, which 
vdSZ calibrated with Galactic bulge PNe. Our new calibration of the relation is
$${\rm log}~D({\rm T_b}) = 3.49 - 0.35~{\rm log}~\theta - 0.32 ~{\rm log}~ F, \eqno(5)$$

based on a fit of the log T$_{\rm b}$ - log R$_{\rm PN}$ relation (R$_{\rm PN}=(\theta$ D)/206265).
We also calibrate the log M$_{\rm ion}$ - log R$_{\rm PN}$ relation as in Maciel \& Pottasch (1980), Z95, and BL01,
and found, by assuming that the filling factor is 0.6, 

$$ {\rm log}~ D({\rm M_{ion}}) = 3.45 - 0.34 ~{\rm log}~\theta - 0.33~ {\rm log}~ F.\eqno(6)$$

Finally, we also recalibrate with the Magellanic Cloud PNe the relation between the surface brightness, I=F/($\pi~\theta^2$), and 
R$_{\rm PN}$, as in SB96, and obtained:

$$ {\rm log}~ D(I) = 3.68 - 0.50 ~{\rm log} ~\theta - 0.25 ~{\rm log} ~F. \eqno(7)$$

The possibility of building a distance scale based on a log I -- log R$_{\rm PN}$ calibration was mentioned by Stanghellini et al. (2002), 
and also used by Jacoby (2002).

All three relations used to derive the distance scales in Eqs. (5), (6), and (7) have high correlation coefficients (R$_{\rm xy} \sim 0.8$). 
In these relations, excluding the bipolar PNe does not change the coefficients by more than 5$\%$, thus their exclusion as calibrators is
irrelevant. 

Using the scales in Eqs. (5), (6), and (7) we have calculated the distances for those Galactic PNe whose individual distances are known either 
through a trigonometric or spectroscopic parallax, or by cluster membership. 
In Table 3 we give the comparison between these newly calibrated scales and the individual distances of PNe. We also give the 
estimates for the LMC and SMC distances, and we infer that the SSV scale is superior to all other scales here recalibrated with the
Magellanic Cloud PNe as well.
We then 
plot in Figure 7 the relative difference between the statistical and individual distances for the scales recalibrated with Magellanic Cloud PNe.
The filled symbols represent the SSV scale, triangles are the distances from the T$_{\rm b}$-R$_{\rm PN}$ relation (Eq. 5), crosses represent the 
log M$_{\rm ion}$ - log R$_{\rm PN}$ scale (Eq. 6),  pentagons are the distances from Eq. (7), based on the
log I -- log R$_{\rm PN}$ relation. We see that the SSV scale is the best possible Galactic statistical distance scale
with the calibrators and comparisons available to date.  Since the log I -- log R$_{\rm PN}$ relation works for bipolar
PNe, it might be used to determine the distance to bipolar PNe instead of the SSV scale.

\section{Conclusion}

The wealth of new data available that describe the physical parameters of Magellanic PNe has allowed us to check and re-calibrate the Shklovsky/Daub/CKS 
statistical distance scale, which is most commonly used in the literature, and provide distances of 645 Galactic PNe following the new distance scale calibration 
(Table 1). To calculate the SSV distance for other PNe, or for the same PNe but using other 
parameters than those in CKS, given  $\theta$ and F, the 5 GHz flux, one can use the following equations:

$$ {\rm log}~D_{\rm SSV} = 3.06 + 0.37~ {\rm log}~ \theta - 0.68~ {\rm log}~F, \tau<2.1 \eqno(8a), $$
$$ {\rm log}~D_{\rm SSV} = 3.79 - 0.6~ {\rm log}~ \theta - 0.2~ {\rm log}~F, \tau>2.1 \eqno(8b). $$

If the 5 GHz flux is not available for the given PN, one can use Eq. (6) in CKS to derive the equivalent 5 GHz flux from the H$\beta$ flux.

In this paper we have used recent data on PNe in the Magellanic Clouds to construct a set of calibrators for which the distances are known 
to high absolute accuracy ($\sim10$\%), and for which the dispersion among the distances is extraordinarily small (a few percent). 
Furthermore, the great distance of these nebulae allows us to establish a distance scale factor that is insensitive to uncertainties 
in distances to Galactic PNe that are drawn from a heterogeneous, nearby (few hundred pc) sample; a local sample has generally been necessary 
given the limited range over which many independent distance methods (notably trigonometric and expansion parallaxes) 
can provide accurate distances.  In addition, we use consistent and reliable means to determine angular sizes (from photometric radii), 
and the H$\beta$ fluxes and extinction constants, derived from \textit{HST} calibrations, are among the most reliable in the literature. 
There has never been a better set of calibrators for statistical distance determinations. For comparison, we selected Galactic PNe where 
the independent distances are the best available (including recently published data), and we evaluated the reliability of various independent 
distance methods by the degree to which they are consistent with our distance scale. 

With this study we show that: 
(1) the distance scale as calibrated from the Magellanic Cloud PNe is very similar to that derived by CKS; 
(2) our revised distance scale agrees superbly with the most accurate distances measured for individual Galactic PNe. We also show that other 
methods of statistical distance determination generally do not yield results that are better than this statistical method;
(3) the distance scale does not work for PNe with bipolar morphology, and we believe this is because progenitors 
of bipolars are often not fully ionized during the course of PN evolution (the log I -- log R$_{\rm PN}$ relation could be used
instead for these PNe);
(4) with the Magellanic Cloud calibration we provide a more robust physical basis for why the Shklovsky/Daub distance scale works, despite wide 
variations in the expected ionized mass; we also show that the recalibration of other distance scales with Magellanic Cloud PNe 
might not work as well as the recalibrated Shklovsky/Daub distance scale;
(5) we point out that the dispersion in the distance scale is an inherent property of the method, and cannot be reduced significantly by using 
better calibrators;
(6) the radiation-bounded sequence for Magellanic Cloud PNe may terminate at higher surface
brightness than previously derived. It seems that the new sequence and the radiation bounded to density bounded transition does not depend
on metallicity very much, as it is the same for the LMC and the SMC PNe; the best available data show that the Magellanic Cloud calibration 
of this sequence is entirely consistent with Galactic PNe.

\acknowledgments

Many thanks to Bruce Balick for scientific discussion on the importance of the 
PN distance scale, and an anonymous referee for giving suggestions that improved the paper.
Letizia Stanghellini is grateful to the Aspen Center for Physics for their hospitality on July 2006, when 
parts of this paper were completed.  


\clearpage

\begin{deluxetable}{lrrrr }
\tablewidth{16truecm}
\tablecaption{Catalog of Galactic PN Distances}
\tablewidth{0pt}
\tablehead{
\colhead{Name}&  
\colhead{$\tau$}&    
\colhead{$\theta$}&   
\colhead{F\tablenotemark{a}}&   
\colhead{D$_{\rm SSV}$}\\

\colhead{}&
\colhead{}&
\colhead{\arcsec}&    
\colhead{}&  

\colhead{[pc]} \\

}

\startdata 
          NGC40 &     3.46 &        18.20&         0.460   &       1249   \\
         NGC246  &   5.31  &     112.00  &      0.248  &        475  \\
         NGC650  &   5.24  &      69.20  &      0.110  &        746  \\
        NGC1360  &   5.82  &     192.00  &      0.222  &        351  \\
        NGC1501  &   4.08  &      25.90  &      0.224  &       1167  \\
        NGC1514  &   4.59  &      50.20  &      0.262  &        760  \\
        NGC1535  &   3.31  &       9.20  &      0.166  &       2305  \\
        NGC2022  &   3.62  &       9.70  &      0.091  &       2518  \\
\enddata
\tablenotetext{a}{This is the 5GHz flux when available, otherwise equivalent 5GHz flux from H$\beta$.}
\end{deluxetable}

\clearpage

\clearpage

\begin{deluxetable}{lcrrrrrr }
\tablewidth{0pt}
\tablecaption{Individual Distances of Galactic PNe}

\tablehead{
\colhead{Name}&
\colhead{D$_{\rm ind}$} &  
\colhead{$\sigma_{\rm D}$} &  
\colhead{$\tau$}&    
\colhead{D$_{\rm CKS}$}&
\colhead{D$_{\rm SSV}$}&
\colhead{Method\tablenotemark{a}}&
\colhead{Ref\tablenotemark{b}}\\

\colhead{}&
\colhead{[pc]} &  
\colhead{[pc]} &  
\colhead{}&    
\colhead{[pc]}&
\colhead{[pc]} &  
\colhead{}&
\colhead{}\\  

}

\startdata 

        A~7     &    676     &    $_{-150}^{+267}$&        6.28  &      216     &    218   &        P    &   Hea07\\
         A~24    &     521   &     $_{-79}^{+112}$&       6.54   &      525     &    530   &        P    &   Hea07\\
         A~31    &     568   &     $_{-90}^{+131}$&       6.97   &      233     &    235   &        P    &   Hea07\\
         BD+30     &    2680   &      810&                   2.1   &     1162     &   3034   &        E    &   HTB93\\
       IC~289    &    2190   &     1630&                  3.81   &     1434     &   1448   &        R    &    KL85\\
      IC~1747    &    2450   &     1150&                  3.12   &     2937     &   2991   &        R   &     KL85\\
      IC~2448    &    1410   &      640&                  3.17   &     3947     &   3984   &        E   &    Pea02\\
     HE~2-131    &     590   &      180&                  2.03   &    1413      &  3666   &       P    &     C99\\
\enddata
\tablenotetext{a} { P: parallax; CM: cluster membership; E: expansion; R: reddening.}
\tablenotetext{b} { ABL00: Alves et al. 2000; CHW03: Chen et al. 2003; C99: Ciardullo et al. 1999; GPP86: 
Gathier et al. 1996; GRM93: Gomez et al. 1993;
HTB93: Hajian et al. 1993; HTB95: Hajian et al. 1995;
HT96: Hajian \& Terzian 1996; Hea07: Harris et al. 2007;  KL85: Kaler \& Lutz 1985; LL68: Liller \& Liller 1968; 
Pea02: Palen et al. 2002: Sea04: Sabbadin et al. 2004.
 }

\end{deluxetable}

\clearpage

\begin{deluxetable}{llcccc }
\tablewidth{18 truecm}
\tablecaption{Comparison of Statistical and Individual Distances}

\tablehead{
\colhead{Ref.}&
\colhead{Method}&
\colhead{R$_{\rm xy}$} &  

\colhead{$<{{\delta{\rm D}} \over {{\rm D}}}>_{\rm Gal}$} &

\colhead{${<D_{\rm stat}>-D_{\rm LMC}} \over {D_{\rm LMC}}$} &

\colhead{${<D_{\rm stat}>-D_{\rm SMC}} \over {D_{\rm SMC}}$} \\

}

\startdata 

SSV&    $\mu\tau$&	0.99&	0.26&	0.01&  0.01\\
&&&&\\

Old scales \\
\hline\\

CKS&	$\mu\tau$&	0.97&	0.32&   0.04&  0.05\\
vdSZ&	log T$_{\rm b}$ -- log R$_{\rm PN}$&    	0.74&	0.86&   0.13&	0.12\\
Z95&	log M$_{\rm ion}$ -- log R$_{\rm PN}$&	0.70&	1.10&   0.19&  0.22\\
BL01&     log M$_{\rm ion}$ -- log R$_{\rm PN}$&  	0.61&   1.35&   0.12&  0.12\\
SB96&   log I -- log R$_{\rm PN}$&    	0.95&   0.46&   0.02&  0.02\\

&&&&\\
LMC/SMC calibrations\\
\hline\\

this paper& log T$_{\rm b}$ -- log R$_{\rm PN}$&	0.82&	0.74&	0.07& 0.02\\
this paper&log M$_{\rm ion}$ -- log R$_{\rm PN}$&	0.80&   0.72&   0.11& 0.08\\
this paper&log I -- log R$_{\rm PN}$&		0.96&   0.36&   0.03& 0.02\\
&&&&\\

\enddata

\end{deluxetable}

\clearpage

\begin{figure}

\plotone{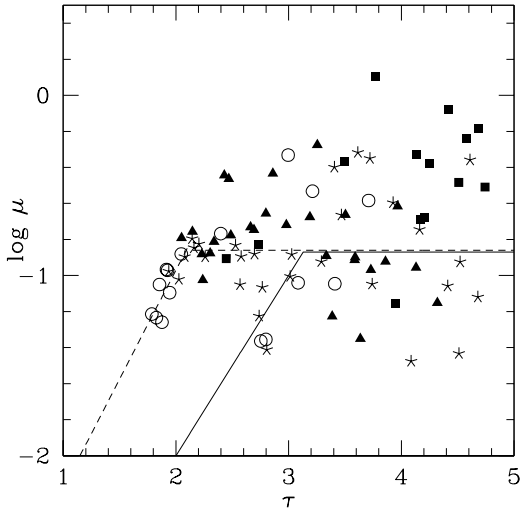}
\vspace{-2cm}
\caption{Plotted are log~$\mu$ versus $\tau$ for the 
sample of LMC PNe observed with the {\it HST}.
Symbols indicate morphology types: Round (open circles), Elliptical 
(asterisks), Bipolar core (triangles) and Bipolar (squares). The thinning
sequence is clearly defined for $\tau<$2.1. The solid line reflect CKS calibration, 
the broken line is the new calibration (SSV). }

\end{figure}

\begin{figure}

\plotone{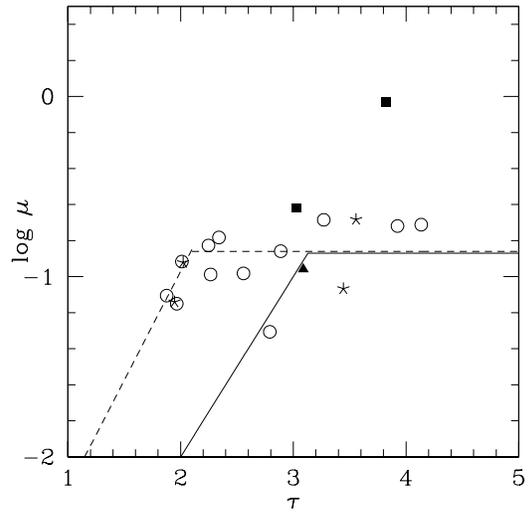}

\caption{Same as in Figure 1, but for the SMC PNe.}

\end{figure}

\begin{figure}

\plotone{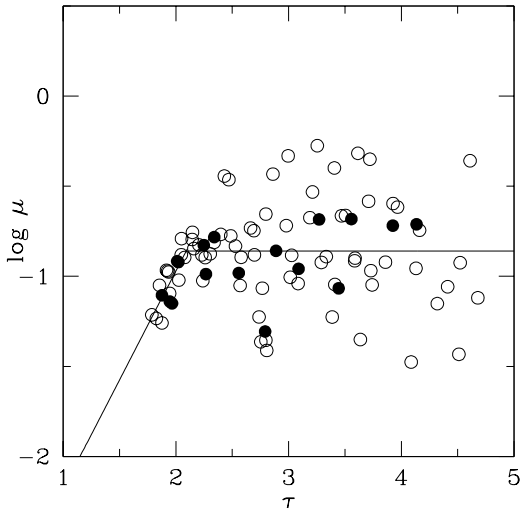}

\caption{LMC (open symbols) and SMC (filled symbols) PN, all morphologies
  except bipolar PN are plotted. Solid line: our new calibration for 
  the Magellanic Cloud PNe.} 

\end{figure}

\begin{figure}

\plotone{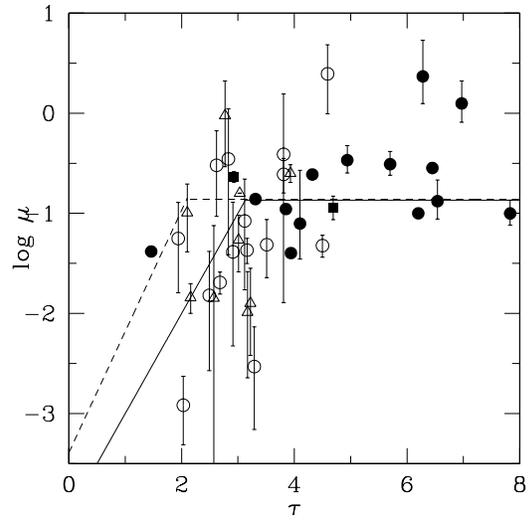}

\caption{The Galactic PNe of known distances plotted on the $\tau$ -- log~$\mu$ plane. Lines as in Figure 3. Symbols denote the
method used for individual distance determination. Filled circles: P; filled squares: CM;
open circles: R; open triangles: E.
}

\end{figure}

\begin{figure}

\plotone{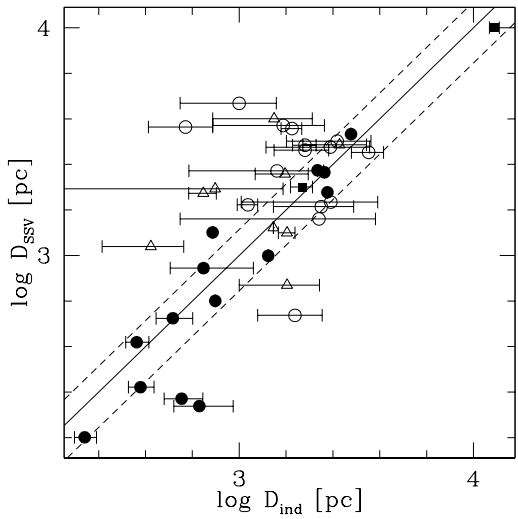}
\vspace{-2cm}
\caption{Comparison of statistical distances form our new calibration (SSV) with individual distances of Table 2. Symbols represent
the individual distance determination method, as in Fig.~5. Solid line: 1:1. Broken lines represent the 30$\%$ differences
between statistical and individual distances.
}

\end{figure}

\begin{figure}

\plotone{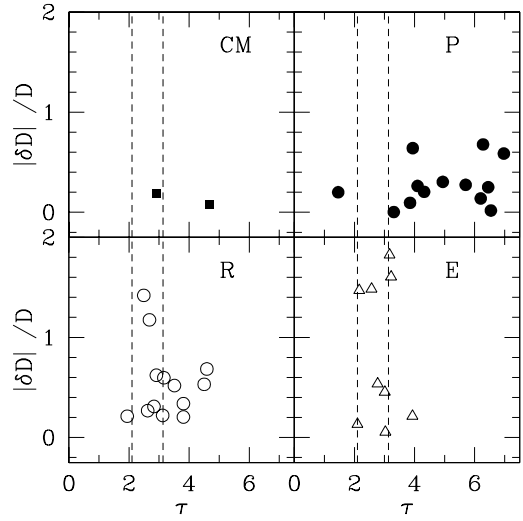}

\caption{Relative differences between SSV statistical and individual PN distances, as a function of $\tau$, separated 
in the panels by individual distance method. Symbols as in Fig.~6. Vertical lines denotes 
the $\tau$ of the thick to thin transition
for the CKS and the SSV scales. 
}

\end{figure}

\begin{figure}

\plotone{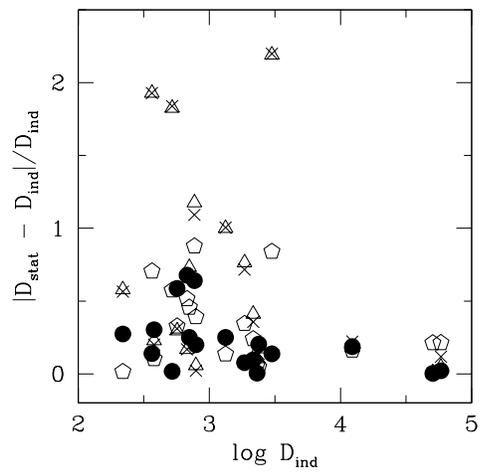}

\caption{Relative differences between statistical and individual PN distances, plotted against the individual distances,
for the Magellanic Cloud PN-calibrated scales of CKS (filled symbols), vdSZ (triangles), BL01 (crosses) and SB96 (pentagons).}

\end{figure}

\end{document}